\documentstyle[multicol,12pt,osa]{revtex}
\begin{document}
\title{On the boundary of the dispersion-managed soliton existence
}

\author{P. M. Lushnikov$^{1,2}$}
\address{
$^1$ Landau Institute for Theoretical Physics, Kosygin St. 2, Moscow,
117334, Russia \\
$^2$Theoretical Division, Los Alamos National
Laboratory, MS-B284, Los Alamos, New Mexico, 87545
\\
E-mail address:  lushnikov@lanl.gov
 }

\maketitle

\begin{abstract}
A breathing  soliton-like structure in dispersion-managed (DM) optical fiber system is studied. It is proven that for negative average dispersion
the breathing  soliton is forbidden provided that a modulus of average
dispersion exceed a threshold which depends on the soliton amplitude.
\end{abstract}

~~~~~~~ {\it PACS numbers:} 05.45.Yv, 42.65.Tg, 42.79.Sz, 42.81.Dp
\\


Propagation of optical pulse in nonlinear media with varying dispersion is both fundamental \cite{zakhmanak1} and important applied
problem \cite{kogelnik1,nakazawa1,smithknox1,gabtur1,kumar1,mamyshev1,mollenauer1}
because a dispersion managed (DM) system, which is a system with periodic
dispersion variation along an optical fiber, is one of the
most prospective candidate for ultrafast high-bit-rate optical communication lines.
Lossless propagation of optical pulse in DM fiber is described by a nonlinear
Schr\"odinger equation (NLS) with periodically varying dispersion $d(z)$:

\begin{equation}
i u_z +   d(z) u_{tt} +  |u|^{2} u =0,  \label{nls1}
\end{equation}
where $u$ is the envelope of the optical pulse, $z$ is the propagation distance and all quantities are
made dimensionless. Consider a two-step periodic dispersion map:
$d(z)=d_0+\tilde d(z)$, where $\tilde d(z)=d_1$ for $0<z+n L<L_1$
and $\tilde d(z)=d_2$ for $L_1<z+n L<L_1+L_2$,  $d_0$ is the path-averaged dispersion, $d_1,d_2$ are the
amplitudes of dispersion variation subject to a condition $d_1L_1+d_2L_2=0$,
$L\equiv L_1+L_2$ is a dispersion compensation period and $n$ is an arbitrary integer number. Eq.
$(\ref{nls1})$  also describes pulse propagation in a fiber with losses compensated for by periodically
placed amplifiers if the distance between amplifiers is much less than $L$.

In a linear regime, in which the nonlinear term in Eq. $(\ref{nls1})$
is negligible, the periodical variation of dispersion is a way to overcome
pulse broadening due to the chromatic dispersion provided that the
residual dispersion $d_0$ is small enough. However in real optical fiber
the nonlinear term in $(\ref{nls1})$ is important because the optical
pulse amplitude should be big enough to get high signal/noise ratio.  One
of the fascinating feature of DM system is the numerical observation of
a space-breathing soliton-like structure, which is called DM soliton, for
both positive and negative residual dispersion $d_0$ \cite{nijhof1}. This
observation is in sharp contrast with the system described by NLS with the
constant dispersion where stable soliton propagation is possible only for
the positive dispersion \cite{zakhint1} because the nonlinearity can
continuously compensate the positive dispersion only. In DM soliton the
balance between the nonlinearity and dispersion  is achieved on average
over the dispersion period $L$ what lift a restriction of the positive
dispersion sign.  Nevertheless it was never proven that DM soliton really
exists because there is a possibility that this is rather a long-lived
quasi-stable breathing pulse which decays on a long distance $z.$ It is
shown here that for negative $d_0$ DM soliton can exist only if $|d_0|$ is
small enough to allow nonlinear compensation of pulse
broadening due to the dispersion over distance $L.$

Eq. $(\ref{nls1})$ can be written in the Hamiltonian form
$\imath u_z=\frac{\delta H}{\delta u^*},$ where the Hamiltonian
\begin{equation} \label{hameq}
H=\int  \left[ d(z)|u_t|^2-\frac{|u|^4}{2}  \right]dt,
\end{equation}
is an integral of motion on each interval of a constant dispersion $d(z)=const.$ Eq. $(\ref{nls1})$ is reduced to usual NLS on such intervals.
At points $z=nL$ and $z=nL+L_1,$ where $n$ is an arbitrary integer number,
the Hamiltonian experiences jumps due to jumps of the dispersion although
the value of $u$ is a continuous function of $z$ in these points.  In
contrast to the Hamiltonian the time-averaged optical power $N=\int
|u|^2dt$ or number of particles in the quantum mechanical
interpretation of NLS (in this interpretation the coordinate $z$ means
some "time" and actual time $t$ has a meaning of "coordinate") is an
integral of motion for all $z.$ Consider z-dependence of the quantity
$A=\int t^2 |u|^2 dt.$ $A/N$ is the average width of a time-distribution of
$u$ or simply $\langle t^2\rangle$ in a quantum mechanical interpretation of
NLS.

Using $(\ref{nls1})$ and integrating by parts one gets for the first $z$ derivative
\begin{eqnarray}\label{At}
A_z=d(z)\int 2 i t  (u u^\ast_t-u^\ast u_t)dt.
\end{eqnarray}
In a similar way after a second differentiation by $z$ one gets
\begin{equation} \label{Att}
A_{zz}=4 d H + 4 d^2 X + \frac{d_z}{d}A_z,
\end{equation}
where $X\equiv \int |u_t|^2 dt.$
It follows from Eq. $(\ref{At})$, which is often called virial theorem (see e.g. Refs.\cite{zakh1972,lush1995}), that  $A_z$ experiences finite jumps corresponding to jumps of a step-wise function $d(z):$
\begin{eqnarray}\label{Azjump1}
  A_z\Big |_{z=L_1+0}=\frac{d_0+d_2}{d_0+d_1}A_z
                              \Big |_{z=L_1-0}\nonumber\\
  A_z\Big |_{z=L+0}=\frac{d_0+d_1}{d_0+d_2}A_z
                              \Big |_{z=L-0} \quad .
\end{eqnarray}

Set $X(z)=X_0+\delta X(z), \quad X(0)\equiv X_0$ then one can integrate
Eq. $(\ref{Att})$ over intervals $(0,L_1),$ $(L_1,L)$:
\begin{eqnarray}\label{Azjump}
  A_z\Big |_{z=L_1-0}=A_z\Big |_{z=0+0}+4\int ^{L_1}_0
     \big [ (d_0+d_1)H_{1}+(d_0+d_1)^2 X\big ] dz \nonumber\\
  A_z\Big |_{z=L-0}=A_z\Big |_{z=L_1+0}+4\int ^{L}_{L_1}
     \big [ (d_0+d_2)H_{2}+(d_0+d_2)^2 X\big ] dz,
\end{eqnarray}
where
\begin{eqnarray}\label{H1H2}
   H_{1}=(d_0+d_1)X_0-Y_0,\nonumber \\
   H_{2}=(d_0+d_2)X_0-Y_0-(d_1-d_2)\delta X\Big |_{z=L_1}
\end{eqnarray}
are the Hamiltonian values on  intervals $(0,L_1), \quad (L_1,L)$ respectively,
$Y(z)\equiv \int \frac{|u|^4}{2}dt, \quad Y_0\equiv Y(0).$
Here the conservation of $H_1$ on interval $(0,L_1)$ is used in derivation
of expression for $H_2$.

The DM soliton solution of Eq. $(\ref{nls1})$ (see Ref. \cite{turitsynsymm1})
is given by $u=\tilde u(z,t) \exp(\imath k z),$ where  $k$ is an arbitrary
real constant and $\tilde u(z+L,t)=\tilde u(z,t)$ is a periodic function
of $z$, $\tilde u(z,t)\big |_{|t|\to \infty}\to 0.$ Thus for DM
soliton $A_z\Big |_{z=L+0}=A_z\Big |_{z=0+0}.$ This condition can be cast
via Eq. $(\ref{Azjump1}),(\ref{Azjump}),(\ref{H1H2})$ into the form:
\begin{eqnarray}\label{2d0}
L(d_1+d_0)\big[2d_0X_0-Y_0+(d_1-d_2)\frac{L_2}{L}\delta X\Big |_{z=L_1}\big ]+\nonumber \\
\int ^{L_1}_0(d_0+d_1)^2 \delta X dz +
\int ^{L}_{L_1}(d_0+d_2)^2 \delta X dz =0.
\end{eqnarray}
Next step is to consider $\delta X(z)$ dependence.
Using $(\ref{nls1})$ and integrating by parts one can get
\begin{equation}\label{Xt}
X_z=4\int\phi_t R_t R^3dt,
\end{equation}
where $u\equiv Re^{\imath \phi},$ $\phi$ and $R$ are real, $R\ge 0$.
Consider an upper bound of $X_z$ which is given by a chain of inequalities:
\begin{equation}\label{Xineq}
4\int\phi_t R_t R^3dt \le
4 \ \max_t(R^2)\int|\phi_t R_t R|dt \le
4 X^{3/2}N^{1/2},
\end{equation}
where the following inequalities are used:
\begin{eqnarray}\label{Xmax}
2\phi_t R_t R\le (\phi_t R)^2+ R_t^2, \nonumber \\
\max_{t}(R^2)\le
\int ^{t}_{-\infty}|(R^2)_{t'}| dt' \le
\int |(R^2)_{t}| dt \le
2\int R|R_{t}| dt \le
2N^{1/2}X^{1/2}
\end{eqnarray}
(in last expression the Cauchy-Schwarz inequality is also used).
Eq. $(\ref{Xt})$ and $(\ref{Xineq})$ can be integrated by $z$ and give (it is assumed
below that  $2X_0^{1/2}N^{1/2} max(L_1,L_2) <1$):
\begin{equation}\label{Xinequp}
X \le
\frac{X_0}{\big (1-2X_0^{1/2}N^{1/2}z \big )^2}
\end{equation}
In a similar way using inequality $X_z\ge -4\int|\phi_t R_t| R^3dt$
following from $(\ref{Xt})$ one can get the lower bond of $X(z):$
\begin{equation}\label{Xineqdown}
X \ge
\frac{X_0}{\big (1+2X_0^{1/2}N^{1/2}z \big )^2}.
\end{equation}

For DM soliton $X(L)=X_0$ and thus it is more convenient to use for $L_1 < z < L$
similar inequalities:
\begin{equation}\label{XinequpdownL2}
\frac{X_0}{\Big (1+2X_0^{1/2}N^{1/2}(L-z) \Big )^2}
\le X \le
\frac{X_0}{\Big (1-2X_0^{1/2}N^{1/2}(L-z) \Big )^2}.
\end{equation}
Eqs. $(\ref{2d0}),(\ref{Xinequp}),(\ref{Xineqdown}),(\ref{XinequpdownL2})$ result in inequality:
\begin{equation}\label{2d0ineqtot}
\begin{array}{cc}
|2d_0X_0-Y_0|\le
  \frac{|d_1-d_2| L_2 X_0}{L}\Big [ \frac{1}{\big (1-2X_0^{1/2}N^{1/2}L_1 \big )^2}-1 \Big ]\\
  + \frac{2  X_0^{3/2}N^{1/2}}{|d_0+d_1| L}\Big [ \frac{(d_0+d_1)^2L_1^2 }{1-2X_0^{1/2}N^{1/2}L_1 }+
      \frac{(d_0+d_2)^2L_2^2}{1-2X_0^{1/2}N^{1/2}L_2}\Big ].
\end{array}
\end{equation}
Eq. $(\ref{2d0ineqtot})$ is the main result of the present paper. Eq. $(\ref{2d0ineqtot})$  is a consequence of initial assumption
that DM soliton exist for given parameters $L_1, L_2, d_0, d_1, d_2$ and integral values $X_0, Y_0, N$ which depend on
$u\big |_{z=0}$ only. Thus DM soliton can exist only if this inequality is fulfilled.

Note that if one assumes uniqueness of DM soliton solution for given $k$
and soliton width then, as shown in Ref. \cite{turitsynsymm1}, $|u|\Big |_{z=0}=|u|\Big |_{z=L_1}$.
In such a case the term $\delta X\Big |_{z=L_1}$ in Eq. $(\ref{2d0})$ vanishes and
instead of $(\ref{2d0ineqtot})$ one can get a stricter inequality.
Here however that possibility is
disregarded for the sake of generality.

To clarify physical consequences of Eq. $(\ref{2d0ineqtot})$
consider the optical pulse with
a typical amplitude $p$ and a typical time-width $t_0.$ Then $N\sim |p|^2
t_0,$ $X_0\sim |p|^2/t_0$ and thus $X_0^{1/2}N^{1/2}L \sim L/Z_{nl},$
where $Z_{nl}=1/|p|^2$ is a characteristic nonlinear length. In a typical
experimental condition a nonlinearity is small: $L/Z_{nl}\ll 1$ and
denominators in $(\ref{2d0ineqtot})$ can be series expanded thus giving
\begin{equation}\label{2d0ineq}
|2d_0X_0-Y_0|\le
  \frac{2 X_0^{3/2}N^{1/2}}{L}\Big [2|d_1-d_2| L_1L_2+
   |d_0+d_1| L_1^2+\frac{(d_0+d_2)^2L_2^2}{|d_0+d_1|}\Big ]
 + O(\frac{d_1 L^3}{t_0 Z_{nl}^3}).
\end{equation}

Provided that $d_0$ is negative both terms in left-hand side of
$(\ref{2d0ineq})$ have the same sign and thus right-hand side
should be greater or equal to $2|d_0|X_0+Y_0.$ Assuming
$d_1\gg |d_0|$ one can get from  $(\ref{2d0ineq})$
the following estimate $(Y_0 \sim t_0/Z_{nl}^2):$
\begin{equation}\label{2d0XYestimate}
\frac{2|d_0|}{t_0 Z_{nl}}+\frac{t_0}{Z_{nl}^2}\stackrel{<}{\sim}
  \frac{4L_1d_1}{Z_{nl}^2 t_0}(1+\frac{L_1}{L}).
\end{equation}
Consider  a strong dispersion management limit $Z_{disp}/L\ll 1,$ where
$Z_{disp}\equiv 2d_1 L_1/t_0^2$ is a typical dispersion length.
This limit implies that an optical pulse experiences strong
oscillation  on each period $L$ due to dispersion.
Then $(\ref{2d0XYestimate})$ reduces to
\begin{equation}\label{2d0Xestimate}
-\frac{d_0}{d_1}\stackrel{<}{\sim} \frac{6 L_1}{Z_{nl}}(1+\frac{L_1}{L}),
\end{equation}
i.e. a nonlinearity (amplitude of the optical pulse) should be strong enough
to allow DM soliton solution existence for a given negative $d_0$.

Eq. $(\ref{2d0ineqtot})$ gives a necessary condition for DM soliton existence but not sufficient.
In other words violation of the inequality $(\ref{2d0ineqtot})$ means that DM soliton is forbidden.
Of course it would be interesting to find to what extent this  necessary existence condition is close to sufficient one.
In general this could be done only if one found DM soliton analytically.
Here one can only mention that there is a qualitative correspondence between
threshold of DM soliton existence following from the analytical condition $(\ref{2d0ineqtot})$ and
from a numerical investigation of DM soliton.
Namely the maximal value of $|d_0|$ $(d_0<0)$ for which DM soliton exist
 grows with increase of the dispersion map strength $L/Z_{disp}$
according to both numerics (see e.g. \cite{berntson1,berntson2}) and the
analytical condition $(\ref{2d0ineq})$.
It
also follows from $(\ref{2d0Xestimate})$ that for asymmetric dispersion
map $L_1\neq L_2$ maximal possible value of $|d_0|$ grows as $L_1$
increase (for fixed $L, Z_{nl}, d_1$) in correspondence with Fig. 3 of
Ref.  \cite{berntson2}.

Eq. $(\ref{2d0ineqtot})$ has also a clear physical meaning in another limit
$\frac{d_0}{d_1}\gg \frac{L}{Z_{nl}},$ $Z_{disp}\gg L$ and $Z_{nl} \gg L$ in which $(\ref{2d0ineqtot})$ reduces to:
\begin{equation}\label{2d0positive}
(2d_0 X_0-Y_0)/Y_0=O(L/Z_{disp})\ll 1.
\end{equation}
Equality  $2d_0 X_0=Y_0$ exactly corresponds to one-soliton solution of NLS with dispersion $d_0$ (see Ref. \cite{zakhint1})
where the dispersion $d_0$ and the nonlinearity continuously balance each other.
Thus in the limit $Z_{disp}\gg L$, which is called a weak dispersion limit, we recover usual NLS describing a path-averaged (over space period L) DM soliton dynamics provided $d_0$ is large enough. A weak dispersion management limit was studied earlier \cite{zakhmanak1,hasegawa1,gablvov1,turmedvedev1}.
Note that an additional condition $\frac{d_0}{d_1}\gg \frac{L}{Z_{nl}}$ allows the amplitude $d_1$ of the dispersion variation still to be much higher
than $d_0$ because one assumes $L\ll Z_{nl}$.

In conclusion the necessary analytical condition $(\ref{2d0ineqtot})$ of DM soliton existence is established. From a physical point of view this condition means that
DM soliton solution can exist only if the nonlinearity is strong enough to compensate the pulse broadening due to the negative value of the average dispersion $d_0.$ Note that estimates in Eqs.
$(\ref{2d0ineq})-(\ref{2d0positive})$ are only given here for a physical interpretation of
the analytical condition $(\ref{2d0ineqtot})$.
So far DM soliton solution was obtained numerically \cite{nakazawa1,smithknox1,berntson1}, by variational \cite{gabtur1} and other perturbative approaches \cite{turmez1,kaup1,lush1}.
These results are in agreement with the condition $(\ref{2d0ineqtot})$. But
analytical proof of DM soliton existence in the parameter region satisfying the condition
$(\ref{2d0ineqtot})$, i.e. the sufficient existence condition, is still an open question.

The author thanks I.R. Gabitov for helpful discussions.

The support was provided by  the US Department
of Energy, under contract W-7405-ENG-36,
RFBR and the program of Russian government support for
leading scientific schools.



\end{document}